\documentclass{aa}
\usepackage{graphicx}
\usepackage{url}
\usepackage{txfonts}
\usepackage{natbib}
\bibpunct{(}{)}{;}{a}{}{,} % to follow the A&A style
\begin{document}
   \title{Experimental \ion{Mg}{\textbf{ix}} photorecombination rate coefficient}

   \author{S. Schippers\inst{1} \and M. Schnell\inst{2} \and C. Brandau\inst{1} \and S. Kieslich\inst{1} \and A.
   M{\"u}ller\inst{1} \and A. Wolf\inst{2}    }

   \offprints{S. Schippers}

   \institute{Institut f{\"u}r Kernphysik, Justus-Liebig-Universit{\"a}t Giessen,
              Leihgesterner Weg 217, D-35392 Giessen, Germany\\
              \email{Stefan.E.Schippers@strz.uni-giessen.de}
         \and
              Max-Planck-Institut f{\"u}r Kernphysik, Saupfercheckweg 1, D-69117 Heidelberg, Germany\\
             }

   \date{Received / Accepted}

\abstract{The rate coefficient for radiative and dielectronic
recombination of berylliumlike magnesium ions was measured with
high resolution at the Heidelberg heavy-ion storage ring TSR. In
the electron-ion collision energy range 0--207 eV resonances due
to $2s \to 2p$ ($\Delta N = 0$) and $2s \to 3l$ ($\Delta N=1$)
core excitations were detected. At low energies below 0.15 eV the
recombination rate coefficient is dominated by strong
$1s^2\,(2s\,2p\,\,^3P)\,7l$ resonances with the strongest one
occuring at an energy of only 21 meV. These resonances decisively
influence the \ion{Mg}{ix} recombination rate coefficient in a low
temperature plasma. The experimentally derived \ion{Mg}{ix}
dielectronic recombination rate coefficient ($\pm 15\%$
systematical uncertainty) is compared with the recommendation by
Mazzotta et al.\ (1998, A\&AS, 133, 403) and the recent
calculations by Gu (2003, ApJ, 590, 1131) and by Colgan et al.\
(2003, A\&A, 412, 597). These results deviate from the
experimental rate coefficient by 130\%, 82\% and 25\%,
respectively, at the temperature where the fractional abundance of
\ion{Mg}{ix} is expected to peak in a photoionized plasma. At this
temperature a theoretical uncertainty in the
$1s^2\,(2s\,2p\,\,^3P)\,7l$ resonance positions of only 100 meV
would translate into an uncertainty of the plasma rate coefficient
of almost a factor 3. This finding emphasizes that an accurate
theoretical calculation of the \ion{Mg}{ix} recombination rate
coefficient from first principles is challenging.

 }

   \maketitle
%
%________________________________________________________________

\section{Introduction}\label{sec:intro}

For the accurate calculation of the ionization equilibrium in astrophysical plasmas,
rate coefficients for the population and depopulation of the various ion charge
states have to be known precisely. To date most of the required rate coefficients
are derived from theoretical calculations, and, hence, experimental benchmarks are
required for testing and improving the theoretical methods. This is especially true
for recombination in photoionized plasmas, which occurs at relatively low plasma
temperatures of only a few electron volts. At such low temperatures the
recombination rate coefficient, depending on the ion under consideration, can
strongly be influenced by the existence of low-energy dielectronic recombination
(DR) resonances that are difficult to theoretically predict with sufficient
accuracy.

Experimentally derived plasma rate coefficients were previously
published for the recombination of \ion{C}{iv}
\citep{Schippers2001c}, \ion{O}{vi} \citep{Boehm2003a},
\ion{Ti}{v} \citep{Schippers1998}, \ion{Ni}{xviii}
\citep{Fogle2003b}, \ion{Ni}{xxvi} \citep{Schippers2000b} and
\ion{Fe}{xviii} -- \ion{Fe}{xxii}
\citep{Savin1997,Savin1999,Savin2002a,Savin2002c,Savin2003a}. Here
the \ion{Mg}{ix} recombination rate coefficient derived from
experimental measurements at a heavy ion storage ring is provided.

The approximate temperature ranges where berylliumlike magnesium
forms in photoionized and in collisionally ionized plasmas can be
obtained from the work of \citet{Kallman2001} who calculated the
fractional abundances of ions in plasmas for a variety of physical
conditions. For photoionized plasmas they find that the fractional
\ion{Mg} {ix} abundance peaks at an ionization parameter of
$\log\zeta = 0.9$ corresponding to a temperature of about 2.8 eV.
The `photoionized zone' may be defined as the range of
temperatures where the fractional abundance of a given ion exceeds
10\% of its peak value. For \ion{Mg}{ix} this corresponds to the
temperature range 2--13~eV. Using the same criterion and the
results of \citet{Kallman2001} for coronal equilibrium the
\ion{Mg}{ix} `collisionally ionized zone' is estimated to extend
over the temperature range 60--170~eV. It should be kept in mind
that these temperature ranges are only indicative. In particular,
they depend on the accuracy of the atomic data base used by
\citet{Kallman2001} and, in case of the photoionization zone, on
the assumed $1/E$ energy dependence of the ionizing radiation.
Nevertheless, the above given temperature ranges will be used in
the discussion below.

From the present measurements it is found that the \ion{Mg}{ix}
recombination rate coefficient in the photoionization zone is
decisively influenced by the presence of a strong DR resonance at
an electron-ion collision energy of 21 meV. Any theoretical
calculation aiming at an accurate \ion{Mg}{ix} rate coefficient
will have to predict this resonance's position with an error of
less than a few meV. Presently available atomic-structure computer
codes are generally not capable of providing results with such an
accuracy ab initio. In the standard theoretical approaches the
situation is often improved by using spectroscopically observed
target energies for the atomic structure
\citep{Colgan2003a,Gu2003b}. As will be shown below, a 100 meV
uncertainty on the theoretical resonance position would translate
into an uncertainty of the \ion{Mg}{ix} DR rate coefficient of a
factor 2.7 at the temperature where the \ion{Mg}{ix} abundance is
expected to peak in photoionization equilibrium.

\section{Experiment}\label{sec:exp}

Various aspects of recombination measurements at the heavy-ion storage ring TSR of
the Max-Planck-Institut f{\"u}r Kernphysik (MPI-K) in Heidelberg, Germany, have been
described by \citet{Kilgus1992}, \citet{Lampert1996}, \citet{Pastuszka1996},
\citet{Mueller1997c}, and \citet{Schippers2000b,Schippers2001c}. For the present
experiment a \ion{Mg}{ix} ion beam with an energy of about 4.2 MeV/u was provided by
the MPI-K linear accelerator facility and injected into the storage ring. There,
over a distance of $\sim$1.5~m, the ion beam was merged with a collinearly moving,
magnetically guided electron beam. The electron beam served two purposes. First, it
acted as a coolant for the ion beam, i.\,e., the ion beam's diameter and momentum
spread were reduced by collisions with the much "colder" electrons \citep[electron
cooling,][]{Poth1990}. Second, the electron cooling device (electron cooler) was
subsequently used as an electron target where the ions recombined with the electrons
and thereby changed their charge state. After adjusting the electron energy to a
well defined value by tuning the electron cooler's cathode voltage appropriately,
recombined ions were separated from the circulating parent-ion beam in the first
bending magnet downstream from the electron cooler. Because of their high velocity
they were efficiently counted with a scintillation detector. \emph{Absolute}
recombination rate coefficients as a function of relative energy between electrons
and ions were derived from the measured count rate by i) subtracting the separately
measured "background" count rate due to electron capture from residual gas molecules
and by ii) normalizing to the number of stored ions and to the electron current
which were both continuously monitored during the measurement. The "merged-beams
technique" was recently reviewed by \citet{Phaneuf1999}.

For the derivation of \emph{absolute} rate coefficients from merged electron-ion beams measurements the
contamination of the ion beam by metastable ions is an issue of concern. In a storage-ring experiment metastable
ions can usually be given enough time (up to a few seconds) to decay before data taking is started. However,
berylliumlike ions with zero nuclear spin like \ion{$^{24}$Mg}{ix} possess an extremely long-lived metastable
$1s^2\,2s\,2p\,\,^3P_0$ state which cannot decay to the $1s^2\,2s^2\,\,^1S_0$ ground state by a one-photon
transition. DR resonances due to the excitation of the $^3P_0$ metastable state were observed in
\emph{single-pass} merged-beams experiments with berylliumlike ions by \citet{Badnell1991a}. Such resonances are
not observed in the present experiment. All measured resonances can be attributed to the excitation of
\ion{Mg}{ix} ground-state ions.

\section{Results}\label{sec:res}

The experimental \ion{Mg}{ix} merged-beams recombination rate coefficient is shown
in Figures \ref{fig:Mg8DN0} -- \ref{fig:Mg8DN1} over different ranges of
electron-ion collision energy. Figure \ref{fig:Mg8DN0} displays all measured
recombination resonances due to $1s^2\,2s^2 \to 1s^2\,2s\,2p$ and $1s^2\,2s^2 \to
1s^2\,2p^2$ ($\Delta N$\,=\,0) core excitations with the exception of the resonances
at very low energies below 0.15~eV. These were measured separately with higher
resolution and are displayed in Figure \ref{fig:Mg8hires}. Finally, Figure
\ref{fig:Mg8DN1} shows at higher energies the DR spectrum for resonances attached to
$1s^2\,2s^2 \to 1s^2\,2s\,3l$ ($\Delta N$\,=\,1) core excitations.

\subsection{$\Delta N$\,=\,0 di- and trielectronic recombination}\label{sec:DN0}

\begin{figure}
 \centering
 \includegraphics[width=\columnwidth]{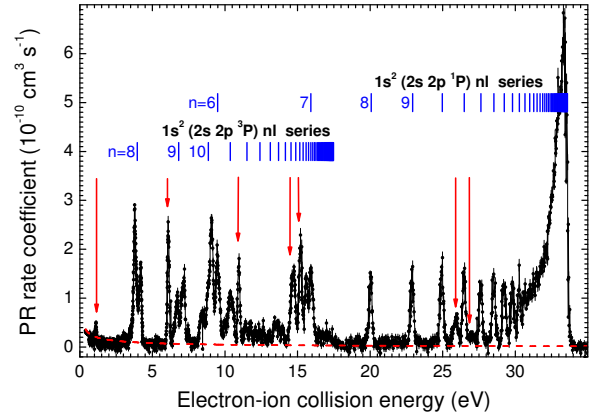}
\caption{\label{fig:Mg8DN0}Measured \ion{Mg}{ix} merged-beams recombination
rate-coefficient in the electron-ion collision energy region of DR
$1s^2\,2s\,2p\,nl$ resonances attached to $1s^2\,2s^2 \to 1s^2\,2s\,2p$ ($\Delta
N$\,=\,0) core excitations. The dashed line is the calculated hydrogenic rate
coefficient for radiative recombination (RR). Positions of some $1s^2\,2p^2\,nl$ TR
resonances are marked by vertical arrows (see text).}
\end{figure}

Most of the measured $\Delta N$\,=\,0 photorecombination resonances in Figure
\ref{fig:Mg8DN0} can be ascribed to $1s^2\,2s^2 \to 1s^2\,(2s\,2p\,\,^3P)$ and
$1s^2\,2s^2 \to 1s^2\,(2s\,2p\,\,^1P)$ excitations. The corresponding Rydberg series
of DR resonances converge to their respective series limits. Their spectroscopic
values are 17.56 and 33.685 eV \citep{Martin1999}. In order to match the
experimental series limits -- obtained by extrapolating the $1s^2\,2s\,2p\,nl$
resonance positions to $n\to\infty$ -- to these values a scaling factor 1.00564 was
applied to the energy scale. This factor is within the experimental uncertainty.

Next to the $1s^2\,2s\,2p\,nl$ DR resonances, further resonances are attached to $2s^2 \to
2p^2$ double core excitations; they appear especially below the $1s^2\,(2s\,2p\,\,^3P)$ series
limit. The radiative decay of these \emph{triply} excited resonance states to below the
\ion{Mg}{ix} ionization limit completes \emph{trielectronic} recombination (TR). The
importance of TR for the photorecombination of berylliumlike ions was only recently discovered
and discussed by \citet{Schnell2003b}. For berylliumlike \ion{Cl}{xiv} they find that over the
plasma temperature range 1--100~eV TR contributes up to 20--40\% to the total recombination
rate coefficient.

The disentanglement of TR and DR contributions to the measured recombination
spectrum requires detailed atomic structure calculations and is beyond the scope of
this paper. Therefore, only obvious TR contributions are marked in Figure
\ref{fig:Mg8DN0}. For a more detailed discussion of TR the reader is referred to the
work of \citet{Schnell2003b}. For the present purpose of deriving the experimental
\ion{Mg}{ix} plasma rate coefficient the question of the origin of  individual
resonances is not relevant. In any theoretical calculation, however, TR has to be
accounted for in order to arrive at an accurate \ion{Mg}{ix} plasma rate
coefficient.

\subsection{Recombination at low energies}\label{sec:lowE}

\begin{figure}
 \centering
 \includegraphics[width=\columnwidth]{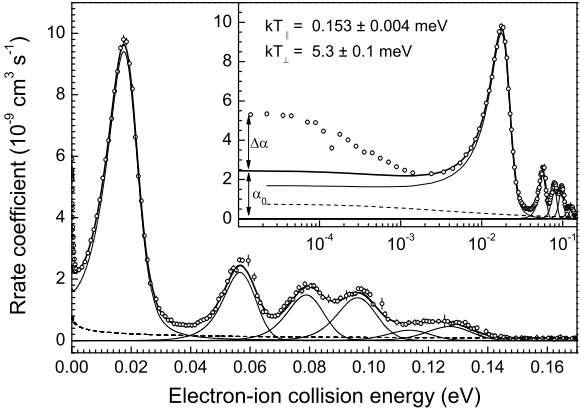}
\caption{\label{fig:Mg8hires}High resolution \ion{Mg}{ix}
merged-beams recombination rate-coefficient in the energy range of
of the $1s^2\,(2s\,2p\,\,^3P)\,7l$ DR resonances close to zero
electron-ion collision energy: Experiment (symbols) and RR and DR
resonance fit (thick full line). Individual contributions to the
fit are 6 DR resonances (thin full lines) and RR (thin dashed
line). Data points below 3~meV were excluded from the fit. The
electron beam temperatures resulting from the fit are indicated in
the inset which shows the same data on a logarithmic energy scale.
It emphasizes the excess recombination rate coefficient $\Delta
\alpha$ at very low energies (see text). The maximum enhancement
is $(\Delta\alpha+\alpha_0)/\alpha_0 \approx 2.3$. Further fit
results are given in table \ref{tab:fit}.}
\end{figure}

Figure \ref{fig:Mg8hires} shows the experimental merged-beams
\ion{Mg}{ix} recombination rate coefficient from very low energies
up to 0.17 eV.  In this energy range, the recombination rate
coefficient is dominated by $1s^2\,(2s\,2p\,\,^3P)\,7l$ DR
resonances. For an accurate determination of the resonance
parameters we fitted 6 DR resonance line profiles to the measured
spectrum as well as a smooth contribution due to radiative
recombination (RR). In the fit the corresponding cross sections
$\sigma(E)$ were convoluted with the experimental electron energy
distribution
\begin{eqnarray}\label{eq:fecool}
\lefteqn{f(E,\hat{E},T_\|,T_\perp) =
\frac{1}{k_\mathrm{B}T_\perp\xi}\,
  \exp\left(-\frac{E-\hat{E}/\xi^2}{k_\mathrm{B}T_\perp}\right)}\hspace{1cm}\\
  &&\times\left[
        \mathrm{erf}\left(\frac{\sqrt{E}+\sqrt{\hat{E}}/\xi^2}{\sqrt{k_\mathrm{B}T_\|}/\xi}\right)
       +\mathrm{erf}\left(\frac{\sqrt{E}-\sqrt{\hat{E}}/\xi^2}{\sqrt{k_\mathrm{B}T_\|}/\xi}\right)
     \right].\nonumber
\end{eqnarray}
It is characterized by the longitudinal and transversal
--- with respect to the electron beam direction --- temperatures $T_\|$ and
$T_\perp$, respectively \citep{Kilgus1992}, and $\xi =
(1-T_\|/T_\perp)^{1/2}$. Furthermore, $k_B$ denotes Boltzmann's
constant and $\mathrm{erf}(x) = 2 \pi^{-1/2}\int_0^x \exp{-t^2}dt$
is the error function.

The convolution of the recombination cross section with the above
experimental electron energy distribution yields the experimental
\emph{merged-beams rate coefficient}
\begin{equation}
\alpha_{MB}(\hat{E}) = \int_0^\infty \!\!\!\!\sqrt{2E/m_e}\, \sigma(E)\,
f(E,\hat{E},T_\|,T_\perp)\, dE\label{eq:alphaMB}
\end{equation}
where $\hat{E}$ and $m_e$ denote the experimental electron-ion
collision energy and the electron rest mass, respectively. For
each DR resonance the cross section was taken to be
\begin{equation}
\sigma^\mathrm{(DR)}(E) = \frac{\overline{\sigma}} {\pi} \frac{E_\mathrm{res}}{E}
\frac{\Gamma/2}{(E-E_\mathrm{res})^2+(\Gamma/2)^2} \label{eq:sigmaDR}.
\end{equation}
The DR line shape is assumed to be a Lorentzian multiplied by a factor
$E_\mathrm{res}/E$ accounting for the $1/E$ dependence of the DR cross section at
low energies \citep{Schippers1998}. For $E_\mathrm{res}\gg\Gamma$ this factor can be
neglected. In the limit $\Gamma\to 0$ the DR-resonance profile becomes a
delta-function and the corresponding merged-beams rate coefficient can be expressed
analytically as
\begin{equation}
  \alpha_{\mathrm{MB}}^{(DR)}(\hat{E}) = \overline{\sigma}
  \sqrt{2E_\mathrm{res}/m_e}\, f(E_\mathrm{res},\hat{E},T_\|,T_\perp)
 \label{eq:alphaDRMB}
\end{equation}
with $f(E_\mathrm{res},\hat{E},T_\|,T_\perp)$ from Equation
\ref{eq:fecool}. For $\Gamma> 0$ the merged-beams rate coefficient
$\alpha_{\mathrm{MB}}^{(DR)}(\hat{E})$ was evaluated numerically.

For the radiative recombination (RR) cross section the semi-classical hydrogenic
formula of \citet{Bethe1957} was used in slightly modified form, i.\,e.,
\begin{equation}\label{eq:sigmaRR}
\sigma^{(RR)}(E) =  2.105\times10^{-22}{\rm cm}^2
\sum_{n_\mathrm{min}}^{n_\mathrm{max}} k_n t_n \, \frac{(Z_\mathrm{eff}^2
\mathcal{R})^2}{n E (Z_\mathrm{eff}^2 \mathcal{R}+n^2 E)}
\end{equation}
and the convolution (Eq.~\ref{eq:alphaMB}) was performed numerically. In equation
\ref{eq:sigmaRR} the letter $\mathcal{R}$ denotes the Rydberg constant. The
effective charge was taken to be $Z_\mathrm{eff} =8$ and the summation over the
principal quantum numbers $n$ was carried out from $n_\mathrm{min}=2$ to
$n_\mathrm{max}=44$. The latter value is determined by field ionization in the
charge-state-analyzing dipole magnet \citep{Schippers2001c}. The constants $t_n$
account for partly filled shells. Here $t_2 = 6/8$ and $t_n = 1$ for $n>2$ were
used. The factors $k_n$ correct for the deviation of the semi-classical cross
sections from the more exact quantum mechanical results and were calculated
following the prescription of \citet{Andersen1990c}. They are monotonically
increasing for increasing $n$, starting from $k_2=0.877$ and approaching unity for
higher $n$.

\begin{table}
\centering
\begin{tabular}{lll}
 \hline
 \noalign{\smallskip}
 $E_\mathrm{res}$ / [meV] & $\Gamma$ / [meV] &
 $\overline{\sigma}$ / [$10^{-18}$ eV cm$^2$]\\
 \noalign{\smallskip}
 \hline
 \noalign{\smallskip}
 ~~20.92(5) & 4.6(1) &   17.93(7)   \\
 ~~60.2(1) &   --   &  ~~2.23(2)   \\
 ~~82.5(1) &   --   &  ~~1.41(2)   \\
 ~~99.9(2) &   --   &  ~~1.29(2)   \\
 117.6(9) &   --   &  ~~0.30(2)   \\
 131.0(4) &   --   &  ~~0.41(2)   \\
 \noalign{\smallskip}
 \hline
\end{tabular}
\caption{\label{tab:fit} Results of fitting individual DR resonances to the measured
recombination rate coefficient at low energies (Fig.\ \ref{fig:Mg8hires}). Each
resonance is characterized by its resonance position $E_\mathrm{res}$, width
$\Gamma$ and strength $\overline{\sigma}$. Numbers in brackets denote the
statistical uncertainties obtained from the fit. If no value for the width is given
a delta-function-like resonance was assumed.}
\end{table}

The resonance parameters that were obtained from the fit are listed in Table~\ref{tab:fit}.
The reduced $\chi^2$ of the fit is 1.24. The electron beam temperatures resulting from the fit
are $k_B T_\| = 0.153 \pm 0.004$~meV and $k_B T_\perp=5.1\pm 0.1$~meV. These values roughly
correspond to what is expected from the electron cooler settings \citep{Pastuszka1996}. Data
points below 3 meV were excluded from the fit. At lower energies the measured rate coefficient
exceeds the fitted one by a factor of up to 2.3 (inset of Figure \ref{fig:Mg8hires}). This
recombination rate enhancement at very low energies is an inherent feature of merged-beams
experiments at electron coolers \citep{Gwinner2000,Heerlein2002,Hoerndl2003a,Wolf2003a}. In
the present context, this recombination rate enhancement can be clearly distinguished from the
normal photorecombination rate coefficient and is excluded from the experimentally derived
plasma rate coefficient of Section \ref{sec:plasma}.

\subsection{$\Delta N$\,=\,1 dielectronic recombination}\label{sec:DN1}

\begin{figure}
 \centering
 \includegraphics[width=\columnwidth]{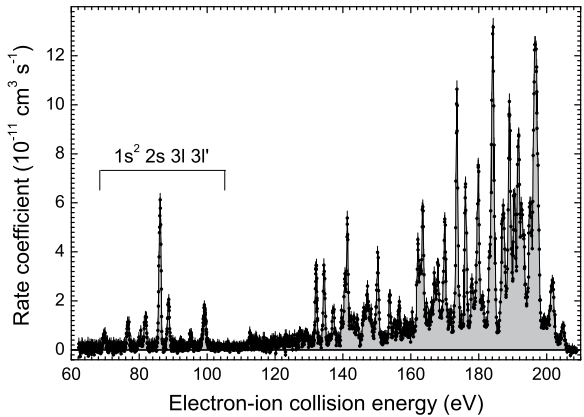}
\caption{\label{fig:Mg8DN1}Measured \ion{Mg}{ix} merged-beams recombination
rate-coefficient in the electron-ion collision energy region of DR
$1s^2\,2s\,3l\,nl'$ resonances attached to $1s^2\,2s^2 \to 1s^2\,2s\,3l$ ($\Delta
N$\,=\,1) core excitations.}
\end{figure}

Resonances attached to $1s^2\,2s^2 \to 1s^2\,2s\,3l$ ($\Delta N=1$) core excitations
occur in the energy range 70-205 eV (Figure \ref{fig:Mg8DN1}). The lowest resonances
of this group are the $1s^2\,2s\,3l\,3l'$ DR resonances extending to energies of up
to about 100~eV. The higher-$n$ manifolds of $1s^2\,2s\,3l\,nl'$ DR resonances
mutually overlap. In contrast to $\Delta N$\,=\,0 DR, the resonance strengths of the
$\Delta N$\,=\,1 DR resonances rapidly decrease with increasing $n$. Therefore, the
resonance strength does not pile up at the series limits. In Figure \ref{fig:Mg8DN1}
two series limits can be discerned at 202.25 and 205.14~eV. In order to match these
values from the NIST atomic spectra data base \citep{Martin1999} for the
$1s^2\,2s\,3d\,\,^1D\,nl$ and the $1s^2\,2s\,3d\,\,^3D\,nl$ series, respectively,
the experimental energy scale was multiplied by a factor 1.0091. This rescaling is
within the experimental uncertainty.

\subsection{Plasma recombination rate coefficient}\label{sec:plasma}

Formally, the recombination rate-coefficient in a plasma is obtained by using an isotropic
Maxwellian distribution function in equation \ref{eq:alphaMB}, i.\,e.,
\begin{equation}
\alpha_\mathrm{plasma}(k_BT) = \frac{4}{\sqrt{2\pi m_e}(k_BT)^{3/2}}\int_0^\infty\!\!\!\!
\sigma(E) E \exp\left(-\frac{E}{k_BT}\right) dE\label{eq:alphaplasma}.
\end{equation}
For plasma temperatures $k_BT \gg k_BT_\perp$ a factor $\sigma(E)\sqrt{2E/m_e}$ in
the integral may be replaced by the experimental merged-beams recombination rate
coefficient. This approach can safely be used for all recombination resonances with
$E_\mathrm{res}\gg k_BT_\perp$ . Here it was applied to all recombination resonances
above 0.2 eV that are shown in Figures \ref{fig:Mg8DN0} and \ref{fig:Mg8DN1}. In
order to exclude any effect of the finite experimental energy spread on the
experimentally derived plasma rate coefficient, Equation \ref{eq:alphaplasma} was
used with the resonance cross section from Equation \ref{eq:sigmaDR} for the
resonances at lower energies. The contribution of the recombination resonances below
0.2~eV (Figure \ref{fig:Mg8hires}) to the plasma recombination rate-coefficient was
evaluated by using the fitted resonance parameters from Table \ref{tab:fit}.

\begin{figure}
 \centering
 \includegraphics[width=\columnwidth]{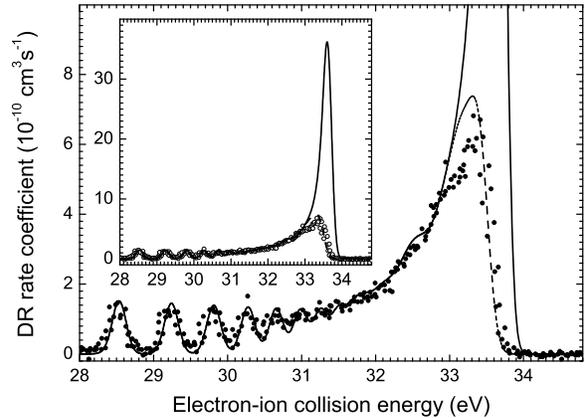}
\caption{\label{fig:Mg8extra} Experimental merged-beams recombination rate
coefficient in the region of high-$n$ $1s^2\,(2s\,2p\,\,^1P)\,nl$ DR resonances
together with the result of an AUTOSTRUCTURE calculation scaled by a factor 0.7. The
dashed line is the theoretical result with account for the experimental field
ionization of high-$n$ Rydberg states using the field-ionization model of
\citet{Schippers2001c}. The full curve is the theoretical result including the full
DR resonance strength up to $n=1000$. The inset shows the same curves on a different
scale for an overview.}
\end{figure}

\begin{table}
\centering
\begin{tabular}{rrr}
 \hline
 \noalign{\smallskip}
$i$ &  $c_i$ / [cm$^{3}$ s$^{-1}$  K$^{3/2}$] & $E_i$ / [eV] \\
 \noalign{\smallskip}
 \hline
 \noalign{\smallskip}
 1&$3.5960\times 10^{-2}$ &  $2.0071 \times 10^{6}$\\
 2&$2.0109\times 10^{-2}$ &  $3.7011 \times 10^{5}$\\
 3&$2.6244\times 10^{-3}$ &  $1.2038 \times 10^{5}$\\
 4&$3.6203\times 10^{-4}$ &  $4.5163 \times 10^{4}$\\
 5&$4.1121\times 10^{-5}$ &  $9.2881 \times 10^{2}$\\
 6&$2.6915\times 10^{-5}$ &  $2.2683 \times 10^{2}$\\
 7&$6.7347\times 10^{-7}$ &  $3.7786 \times 10^{1}$\\
 \noalign{\smallskip}
 \hline
\end{tabular}
\caption{\label{tab:plasma} Parameters for the fit of equation \ref{eq:DRfit} to the
experimentally derived \ion{Mg}{ix} DR+TR rate coefficient in a plasma. In the
temperature ranges 25--350 K and 350--$9\times 10^8$ K the fit deviates less than
2\% and 1\%, respectively, from the experimentally derived result. The systematic
uncertainty of the experimentally derived absolute recombination rate coefficient is
$\pm 15\%$ \citep{Lampert1996}.}
\end{table}

For the derivation of the experimental \ion{Mg}{ix} recombination rate coefficient
one has to be aware of the fact that loosely bound high-$n$ Rydberg states are
easily field ionized in our experimental setup. This effect was modelled in
considerable detail by \citet{Schippers2001c}. Generally, it is more significant for
lower charged ions. As shown in Figure \ref{fig:Mg8extra} a theoretical calculation
of the \ion{Mg}{ix} DR rate coefficient using the AUTOSTRUCTURE code of
\citet{Badnell1986} can only be reconciled with the experimental data if
field-ionization is taken into account.

In the present case the effect of field-ionization on the measured DR spectrum can
be approximately described as a cutoff of high-$n$ Rydberg resonances with $n
\gtrsim 44$. In order to account for the unmeasured DR resonance strength the DR
spectrum for $n<44$ is extrapolated to higher $n$ by scaling the result of the
AUTOSTRUCTURE calculation to the experimental spectrum in the energy range
28.0--32.9~eV (scaling factor 0.7, Figure \ref{fig:Mg8extra}) and by replacing the
measured merged-beams rate coefficient by the scaled calculated value in the energy
range 32.9--34.1~eV.

\begin{figure}
 \centering
 \includegraphics[width=\columnwidth]{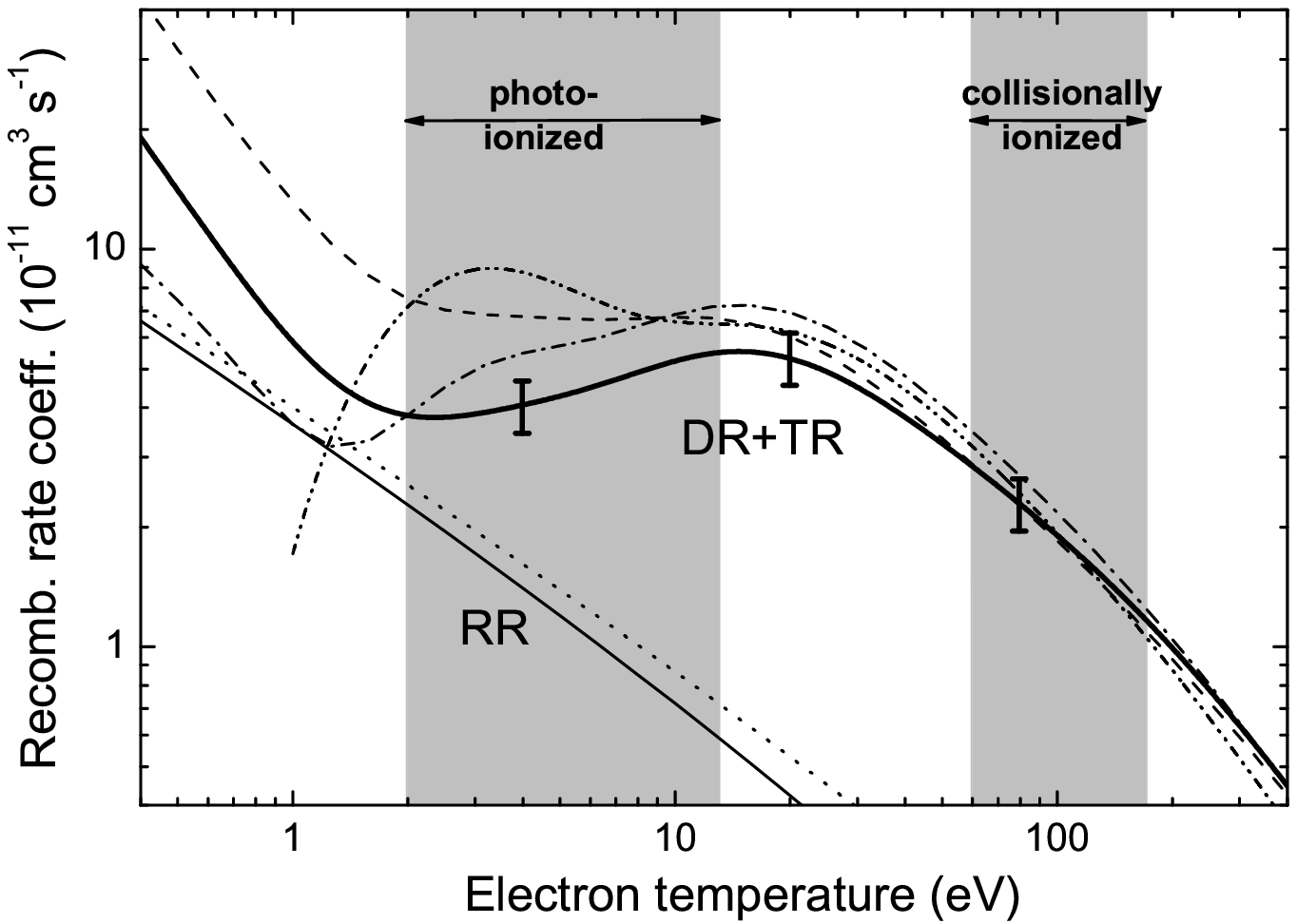}
\caption{\label{fig:Mg8plasma} Comparison of the present
experimentally derived \ion{Mg}{ix} DR rate coefficient (thick
full line) with the recommendation of \citet[][ dash-dot-dotted
line]{Mazzotta1998} and the theoretical results of \citet[][
dashed line]{Gu2003b} and \citet[][ dash-dotted
line]{Colgan2003a}. The error bars denote the $\pm$15\%
experimental uncertainty of the absolute rate coefficient. The
thin full line is the RR rate coefficient calculated (eq.\
\ref{eq:alphaplasma}) with the RR cross section from equation
\ref{eq:sigmaRR} with $n_\mathrm{max}=1000$ and the remaining RR
parameters from section \ref{sec:lowE}. The dotted line is the
theoretical RR rate coefficient of \citet{Gu2003c}. The
temperature ranges (see section \ref{sec:intro}) where
\ion{Mg}{ix} exists in photoionized and in collisionally ionized
plasmas, respectively, are highlighted.}
\end{figure}

In principle, such a correction also has to be made for all other series of Rydberg
resonances. For $\Delta N$\,=\,0 DR such other series are relatively weak (Figure
\ref{fig:Mg8DN0}) and their cutoff by field-ionization does not lead to a significant error
when considering the overall $\pm 15\%$ experimental uncertainty of the absolute rate
coefficient scale \citep{Lampert1996}. Regarding $\Delta N$\,=\,1 DR, the resonance strength,
as already mentioned, decreases strongly with increasing $n$, leading again to a negligible
effect of the experimental field-ionization on the plasma rate coefficient.

Our experimentally derived \ion{Mg}{ix} DR+TR recombination rate coefficient in a
plasma --- including the extrapolation of the $1s^2\,(2s\,2p\,\,^1P)\,nl$ resonance
strength to $n=1000$ (cf.\ Figure \ref{fig:Mg8extra}) --- is shown in Figure
\ref{fig:Mg8plasma} as a thick full line. For the convenient use of our result in
plasma modelling codes  the following functional dependence
--- customarily used for DR rate coefficients --- was fitted to our experimentally derived
curve:
\begin{equation}
\alpha^\mathrm{(DR)}_\mathrm{fit} = \frac{1}{T^{3/2}}\sum_i c_i
\exp\left(-\frac{E_i}{k_BT}\right).\label{eq:DRfit}
\end{equation}
During the fitting procedure the coefficients $c_i$ and $E_i$ were varied. The fit
results are listed in Table \ref{tab:plasma}.

In the fit to the low-energy experimental merged-beams
recombination rate coefficient (Figure \ref{fig:Mg8hires}) RR was
also included (with $T_\|$ and $T_\perp$ as the only free fit
parameters and all other parameters kept fixed). The same RR cross
section was used to draw the RR contribution in Figure
\ref{fig:Mg8DN0}. This shows that Equation \ref{eq:sigmaRR} for
the RR cross section is consistent with our experimental data.
Therefore, Equation \ref{eq:sigmaRR}, in conjunction with Equation
\ref{eq:alphaplasma}, can also be used to derive the experimental
\ion{Mg}{ix} RR rate coefficient in a plasma.

To this end, the RR cross section parameters were taken from
section \ref{sec:lowE}. In order to make up for field-ionization
effects the plasma RR rate coefficient in Figure
\ref{fig:Mg8plasma} (thin full line) was calculated with
$n_\mathrm{max}=1000$ instead of $n_\mathrm{max}=44$ as in Figure
\ref{fig:Mg8hires}. For convenient further use our \ion {Mg}{ix}
RR rate coefficient in a plasma was fitted by a formula introduced
by \citet{Verner1996a}:
\begin{equation}
\alpha^\mathrm{(RR)}_\mathrm{fit} = A \left[
\sqrt{\frac{T}{T_0}}\left(1+\sqrt{\frac{T}{T_0}}\right)^{1-b}
\left(1+\sqrt{\frac{T}{T_1}}\right)^{1+b}\right]^{-1}.\label{eq:RRfit}
\end{equation}
The parameter values that were obtained from the fit are
$A=5.7414\times10^{-10}$~cm$^3$~s$^{-1}$, $b = 0.67692$, $T_0 =
206.15$~K, and $T_1 = 7.6200\times 10^{6}$~K. Over the temperature
ranges $1 - 2.9\times 10^6$~K and $2.9\times 10^6 - 10^8$ K the
fit is accurate to more than 2\% and 3\%, respectively.

In Figure \ref{fig:Mg8plasma} the present experimentally derived RR rate coefficient is
compared with the the recent theoretical result of \citet[][dotted line]{Gu2003c}. The
difference is less than 17\% over the temperature range where \ion{Mg}{ix} exists in a
photoionized plasma.

The total (unified) \ion{Mg}{ix} recombination rate coefficient is readily obtained as the sum
of our DR+TR and RR rate coefficients. It should be noted that this result does not depend on
the existence of possible quantum mechanical interferences between RR and DR, since
interference has already been neglected during the decomposition of the measured spectrum into
DR and RR (Figure \ref{fig:Mg8hires}). In principle, interference between RR and DR can lead
to asymmetric recombination resonance line shapes \citep[see e.\,g.][]{Schippers2002a}.  No
evidence for such effects is found in our experimental \ion{Mg}{ix} recombination spectra. The
systematic uncertainty of the experimentally derived total \ion{Mg}{ix} recombination rate
coefficient is $\pm 15\%$ \citep{Lampert1996}.

\section{Discussion and Conclusions}\label{sec:dis}

In Figure \ref{fig:Mg8plasma} the present experimentally derived
DR+TR rate coefficient is compared with the recommendation of
\citet{Mazzotta1998} and with the recent calculations of
\citet{Gu2003b} and \citet{Colgan2003a}. The recommended
\ion{Mg}{ix} DR rate coefficient of \citet{Mazzotta1998} deviates
by up to 130\% from the experimental result in the temperature
range 2--13~eV where \ion{Mg}{ix} exists in photoionized plasmas.
This rather large discrepancy questions the usefulness of the
recommended \ion{Mg}{ix} DR rate coefficient for the modeling of
photoionized plasmas. In the temperature range 60-170~eV
(collisionally ionized zone) the recommended \ion{Mg}{ix} DR rate
coefficient of \citet{Mazzotta1998} agrees within 11\%, i.\,e.,
within the 15\% experimental uncertainty, with the experimentally
derived one.

In this temperature range the theoretical rate coefficient of
\citet{Gu2003b} shows very good agreement with the experimentally
derived rate coefficient. The difference between the two curves in
this range is less than 7\%. At lower temperatures, however, the
discrepancy is up to 99\% at 2~eV. At these low temperatures where
\ion{Mg}{ix} exists in photoionized plasmas, the theoretical
result of \citet{Colgan2003a} agrees better with the experimental
rate coefficient. The deviation is less than 35\% for $k_BT
> 2$~eV, less than 30\% for $k_BT > 15$~eV, and less than 15\%
for $k_BT > 100$~eV. At the temperature 2.8 eV  where the
fractional abundance of \ion{Mg}{ix} is expected to peak in a
photoionized plasma \citep{Kallman2001} the results of
\citet{Gu2003b} and \citet{Colgan2003a} deviate from the
experimental rate coefficient by 82\% and 25\%, respectively.

\begin{figure}
 \centering
 \includegraphics[width=\columnwidth]{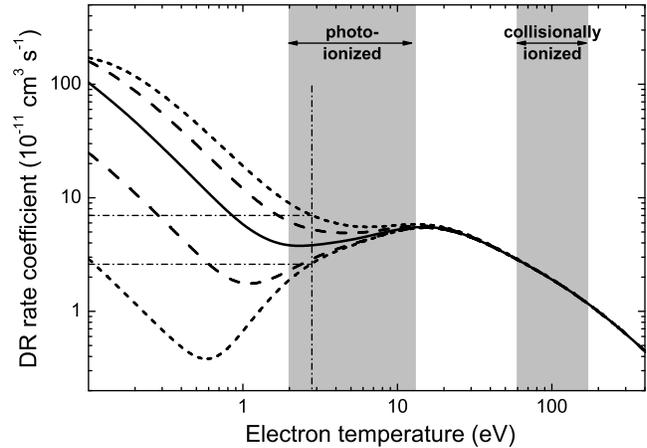}
\caption{\label{fig:Mg8shift}Impact of a shift of the low energy
$1s^2\,(2s\,2p\,\,^3P)\,7l$ resonances displayed in Figure
\ref{fig:Mg8hires} on the \ion{Mg}{ix} DR plasma rate-coefficient.
The full curve is the experimental result, the long-dashed curves
correspond to resonance shifts of $\Delta E_\mathrm{res} = \pm
50$~meV, and the short-dashed  curves to a shift of $\Delta
E_\mathrm{res} = \pm 100$~meV. Thereby, the lower curves
correspond to resonance shifts to lower energies. The vertical
dash-dotted line marks the temperature 2.8~eV where the fractional
abundance of \ion{Mg}{ix} peaks in a photoionized gas
\citep{Kallman2001}. The horizontal dash-dotted lines indicate the
factor 2.7 uncertainty of the plasma rate coefficient at this
temperature that would be introduced by a $\pm 100$~meV
uncertainty of the low energy resonance positions.}
\end{figure}

The rather large low-temperature deviation of the theoretical rate
coefficient of \citet{Gu2003b} from the experimentally derived
curve is most probably due to an inaccurate calculation of the DR
resonance energies below 0.15~eV (Figure \ref{fig:Mg8hires}). At
these very low energies the DR rate coefficient is very sensitive
to slight variations of resonance positions. This is highlighted
in Figure \ref{fig:Mg8shift} that displays the effect of
hypothetical shifts of the low-energy resonance positions by $\pm
50$ meV and $\pm 100$~meV with respect to the tabulated values
(Table \ref{tab:fit}) on the \ion{Mg}{ix} DR rate coefficient.
Obviously, a $\pm 100$~meV uncertainty of the low-energy
$1s^2\,(2s\,2p\,\,^3P)\,7l$ resonance positions translates into an
uncertainty of a factor 2.7 of the plasma rate coefficient in the
temperature range where the fractional abundance of \ion{Mg}{ix}
peaks in a photoionized gas.

These findings demonstrate that an accurate calculation of DR
rates for ions that exhibit DR resonances close to zero energy is
challenging. In case of more complex ions this task is certainly
beyond the capabilities of present-day atomic structure codes as
is exemplified by a recent combined theoretical and experimental
DR study of argon-like Sc$^{3+}$ \citep{Schippers2002a}.
Recombination experiments at heavy-ion storage rings are certainly
required for guiding the future development of the theoretical
methods, especially in the case of low temperature DR rate
coefficients for complex ions.

\begin{acknowledgements}
      We thank D. W. Savin for helpful discussions and
      the MPI-K accelerator crew, in particular R. Repnow and M. Grieser, for their excellent support.
      This work was partly funded by Deut\-sche For\-schungs\-ge\-mein\-schaft under contract Mu 1068/8.
\end{acknowledgements}

%\bibliographystyle{aa}
%\bibliography{e:/tex/schippers}

\end{document}